\documentclass[11pt]{article} 
\newcommand{\be}{\begin{equation}}
\newcommand{\ee}{\end{equation}}
\newcommand{\bea}{\begin{eqnarray}}
\newcommand{\eea}{\end{eqnarray}}
\newcommand{\sn}{{\rm sn}}
\newcommand{\ds}{{\rm ds}}
\newcommand{\cs}{{\rm cs}}
\newcommand{\ns}{{\rm ns}}
\newcommand{\dn}{{\rm dn}}
\newcommand{\cn}{{\rm cn}}
\newcommand{\sech}{{\rm sech}}

\begin{document}
\vspace{.5in} 
\begin{center} 
{\LARGE{\bf Superposition of Elliptic Functions as Solutions For a 
Large Number of  Nonlinear Equations}}
\end{center} 

\vspace{.3in}
\begin{center} 
{\LARGE{\bf Avinash Khare}} \\ 
{Raja Ramanna Fellow,}
{Indian Institute of Science Education and Research (IISER), Pune, India
411021}
\end{center} 

\begin{center} 
{\LARGE{\bf Avadh Saxena}} \\ 
{Theoretical Division and Center for Nonlinear Studies, Los
Alamos National Laboratory, Los Alamos, NM 87545, USA}
\end{center} 

\vspace{.9in}
{\bf {Abstract:}}  

For a large number of nonlinear equations, both discrete and continuum, we 
demonstrate a kind of linear superposition.  We show that whenever a nonlinear 
equation admits solutions in terms of both Jacobi elliptic functions $\cn(x,m)$ and 
$\dn(x,m)$ with modulus $m$, then it also admits solutions in 
terms of their sum as well as difference. We have checked this in the case of
several nonlinear equations such as the nonlinear Schr\"odinger equation, MKdV, 
a mixed KdV-MKdV system, a mixed quadratic-cubic nonlinear Schr\"odinger equation,
the Ablowitz-Ladik equation, the saturable nonlinear Schr\"odinger equation, $\lambda
\phi^4$, the discrete MKdV as well as for several coupled field equations. Further, for 
a large number of nonlinear equations, we show that whenever a nonlinear equation 
admits a periodic solution in terms of $\dn^2(x,m)$, it also admits solutions in terms of 
$\dn^2(x,m) \pm \sqrt{m} \cn(x,m) \dn(x,m)$, even though  $\cn(x,m) \dn(x,m)$ is not a
 solution of these nonlinear equations. Finally, we also obtain  superposed
solutions of various forms for several coupled nonlinear equations. \\

\noindent {\bf Date of resubmission}: February 7, 2014

\newpage 
  
\section{Introduction} 

Nonlinear equations are playing an increasingly important role in several areas 
of science in general and physics in particular. One of the major problem with these 
equations is the lack of a superposition principle. In this context it is worth recalling 
that the linear superposition principle is one of the hallmarks of linear theories
which does not hold good in nonlinear theories because of the nonlinear
term(s). Thus, even if two solutions are known for a nonlinear theory, their
superposition is in general not a solution of that nonlinear theory. The purpose of 
this paper is to point out a novel kind of superposition which seems to hold good for 
a large number of nonlinear equations, both discrete and continuum. In particular,
there are several nonlinear field equations, discrete as well as continuum \cite{rev, 
flach, PGK_book}, which are known to admit periodic solutions in terms of Jacobi 
elliptic functions (JEF) $\cn(x,m)$ and $\dn(x,m)$, where $m$ denotes the modulus 
of the elliptic function \cite{as}. Many of these solutions have found wide application 
in several areas of physics \cite{1,2,rajaraman}.  Our goal is to show, through a large 
number of examples, that a kind of novel linear superposition seems to hold good in 
these cases.  We might add here that we do not have a rigorous proof for such a 
superposition but we have examined a large number of examples, which without 
exception, seem to support this conjecture.  In particular, we examine a number of 
nonlinear equations, both continuum and discrete, both integrable and nonintegrable,  
which admit periodic solutions in terms of $\cn(x,m)$ and $\dn(x,m)$ functions and 
show that in all these cases, without exception,  $\dn(x,m) \pm \sqrt{m} \cn(x,m)$ are 
also exact periodic solutions. 

The continuum nonlinear equations that we have studied are the nonlinear 
Schr\"odinger equation (NLSE), quadratic-cubic NLSE \cite{qcnls1, qcnls2}, MKdV 
\cite{1, 2, rajaraman}, mixed KdV-MKdV system, $\lambda \phi^4$ field theory 
\cite{1,2, rajaraman}, etc. On the other hand, the discrete nonlinear equations that we 
have examined are the Ablowitz-Ladik equation \cite{al, takeno}, saturable discrete NLSE 
\cite{ak4}, discrete MKdV, discrete $\lambda \phi^4$ field theory \cite{ak7}, discrete 
cubic-quintic model, etc.  Amongst these, NLSE, MKdV, Ablowitz-Ladik and discrete 
MKdV are the integrable models while the rest are not integrable.  In addition, we have 
studied several coupled nonlinear equations, e.g.  coupled $\phi^4$ \cite{ak1}, coupled 
NLS-MKdV system, coupled KdV-quadratic NLS, coupled NLS (including the Manakov 
system \cite{man}), etc. and find that they also admit such superposed solutions. 

Further, we also examine a number of continuum field theories like KdV \cite{1,2, 
rajaraman}, quadratic NLS and $\phi^3$ field theory \cite{phi3, tachyon} which admit 
$\dn^2(x,m)$ as a periodic solution and show that all these models also admit periodic 
solutions of the form $\dn^2(x,m) \pm \sqrt{m} \cn(x,m)\dn(x,m)$ even though $\cn(x,m) 
\dn(x,m)$ is not a solution of such models. While this cannot be viewed as a linear 
superposition of two solutions since $\cn(x,m) \dn(x,m)$ is not a solution of these models, 
it is rather remarkable that such solutions exist, without exception, in a number of 
continuum field theory models, both integrable and nonintegrable, that we have examined. 

We have also considered several coupled field theories in which while one field 
admits a periodic solution of the form $\dn^2(x,m)$, the other field either admits a 
periodic solution of the form $\dn(x,m)$ or $\cn(x,m)$ and in all such cases, without fail, 
we find that the coupled model also admits periodic solutions of the form $\dn^2(x,m) 
\pm \sqrt{m} \cn(x,m) \dn(x,m)$ in one field and $\dn(x,m) \pm \sqrt{m}\cn(x,m)$ type solutions 
in the other field.  We might add here that so far as we are aware of, the solutions obtained 
in this paper are new  and were not known previously. Additionally, the superposed 
solutions that we have obtained are not connected to the known solutions by any kind 
of Landen transformation \cite{as}.

The paper is organized as follows. In Sec. II we discuss several continuum models 
which admit $\cn(x,m)$ as well as $\dn(x,m)$ as periodic solutions and show that such 
models also admit periodic solutions of the form $\dn(x,m) \pm \sqrt{m} \cn(x,m)$.  In 
Sec. III We  discuss a few coupled continuum models in which both the fields are 
known to admit $\cn(x,m)$ and $\dn(x,m)$ as their exact solution and show that such 
coupled models also admit superposed solutions of the form $\dn(x,m) \pm \sqrt{m} 
\cn(x,m)$ in both the fields.  In Sec. IV we discuss several discrete nonlinear equations 
which are known to admit $\cn(x,m)$ and $\dn(x,m)$ as periodic solutions and show 
that all of them also admit $\dn(x,m) \pm \sqrt{m} \cn(x,m)$ as solutions. In Sec. V we 
discuss a few continuum models which admit $\dn^2(x,m)$ as a periodic solution and 
show that such models also admit $\dn^2(x,m) \pm \sqrt{m} \cn(x,m) \dn(x,m)$ as
periodic solutions even though $\cn(x,m) \dn(x,m)$ is not a solution of these models. 
In Sec. VI we discuss few coupled continuum models in which both the fields
are known to admit $\dn^2$ as their exact solution and show that they also admit
solutions of the form $\dn^2 \pm \sqrt{m} \cn \dn$ in both the fields. In Sec. VII we 
discuss a few coupled continuum field theories in which one of the field admits 
$\cn(x,m)$ as well as $\dn(x,m)$ as an exact solution while the other field has 
$\dn^2(x,m)$ as an exact solution. We show that such models also admit $\dn(x,m) 
\pm \sqrt{m}\cn(x,m)$ as solution in the first field and $\dn^2(x,m) \pm \sqrt{m} \cn(x,m) 
\dn(x,m)$ as solution in the second field.  Some preliminary results have appeared 
previously \cite{ak2}.  We summarize our main conclusions in Sec. VIII where we also 
discuss possible reasons why the linear superposition of $\dn(x,m)$ and $\cn(x,m)$ is 
also a solution of models which admit $\dn(x,m)$ and $\cn(x,m)$ as solutions.
  
\section{$\dn \pm \sqrt{m} \cn$ as Exact Solutions of Continuum 
Nonlinear Equations}

In this section, we discuss six continuum models, all of which admit periodic solutions 
in terms of Jacobi elliptic functions (JEF) $\dn(x,m)$ as well as $\cn(x,m)$, and show 
that, in all these cases, $\dn(x,m) \pm \sqrt{m}\cn(x,m)$ are also exact solutions. Hence 
forth, for the sake of brevity, we will omit the arguments $(x,m)$ of JEF in the text. 

\subsection{NLS Equation}

We start with the nonlinear Schr\"odinger (NLS) equation \cite{rev, flach} 
\be\label{1}
iu_t+u_{xx}+g|u|^2 u  =0\,,
\ee
which is a well known integrable model. It has found applications in several
branches of physics \cite{rev, flach}.
It is well known that one of the exact moving periodic 
solution to this equation is
\be\label{2}
u = A \dn[\beta(x-vt+\delta_1),m]\exp[-i(\omega t-kx+\delta)]\,,
\ee
provided
\be\label{3}
gA^2 = 2\beta^2\,,~~\omega = k^2 -(2-m) \beta^2\,,~~v=2k\,.
\ee
Here $\delta,\delta_1$ are two arbitrary constants arising due to translational
invariance. In fact this is true for all the models discussed in this paper 
and hence we will not mention about $\delta,\delta_1$ any more in this paper. 

Another exact solution to NLS Eq. (\ref{1}) is
\be\label{4}
u = A \sqrt{m} \cn[\beta(x-vt+\delta_1),m]\exp[-i(\omega t-kx+\delta)]
\ee
provided
\be\label{5}
gA^2 = 2\beta^2\,,~~\omega = k^2 -(2m-1)\beta^2\,,~~v=2k\,.
\ee

Remarkably, even a linear superposition of the two, i.e.
\be\label{6}
u = \bigg ( \frac{A}{2} \dn[\beta(x-vt+\delta_1),m]
+ \frac{B}{2} \sqrt{m} \cn[\beta(x-vt+\delta_1),m] \bigg )
\exp[-i(\omega t-kx+\delta)]\,,
\ee
is an exact solution to the NLS Eq. (\ref{1}) provided
\be\label{7}
B = \pm A\,,~~gA^2 = 2 \beta^2\,,~~\omega-k^2 =-(1/2)(1+m)\beta^2\,,
~~v=2k\,.
\ee

It is worth noting that the frequency $\omega$ associated with the three solutions 
[i.e. $\cn, \dn$, and ($\dn \pm \sqrt{m} \cn$)] is different except at $m=1$. 
We thus have two new periodic solutions of NLSE depending on whether
$B=A$ or $B=-A$. Few remarks are in order here which are in fact valid for all
the solutions (both continuum and discrete) discussed in Secs. II, III and IV.
Therefore, we shall not repeat these remarks while discussing various 
solutions in these three sections.

\begin{enumerate}

\item All the models in these three sections admit $\cn, \dn$ as well 
as $\dn \pm \sqrt{m} \cn$ 
as solutions. It is insightful to note that both the solutions $\dn +\sqrt{m} \cn$
and $\dn-\sqrt{m} \cn$ exist for the same values of the parameters. It will
be interesting to know the region of stability of these two solutions vis a 
vis those of $\dn$ and $\cn$ solutions. 

\item In the limit $m=1$, all three solutions $\dn$, $\cn$ as well 
$\dn+\sqrt{m}\cn$ go over to the well known pulse (i.e. $\sech$) solution
while $\dn-\sqrt{m} \cn$ goes over to the vacuum solution. 

\item In all the continuum models discussed in this section, 
the factors of $2-m$ or $2m-1$ which 
appear in the $\dn$ and $\cn$ solutions, get
replaced by the factor of $(1+m)/2$ in the $\dn \pm \sqrt{m}\cn$ 
solutions. 

\item On the other hand, in the discrete models discussed in 
Sec. IV, the factors of $\dn(\beta,m)$ or $\cn(\beta,m)$ appearing in 
$\dn$ and $\cn$ solutions, get replaced by the factor of 
$[\dn(\beta,m)+\cn(\beta,m)]/2$ in the $\dn \pm \sqrt{m} \cn$ solutions. 

\end{enumerate}

\subsection{MKdV Equation}

We now show that the celebrated MKdV equation \cite{1,2} 
\be\label{8}
u_t+u_{xxx}+gu^2 u_{x} =0\,,
\ee
which is a well known integrable equation and has found applications in several
areas \cite{1,2}, also admits such  superposed solutions. 

It is well known that one of the exact solution to the 
MKdV Eq. (\ref{8}) is
\be\label{9}
u = A \dn[\beta(x-vt+\delta_1),m]\,,
\ee
provided
\be\label{10}
gA^2 = 6\beta^2\,,~~v=(2-m) \beta^2\,.
\ee
Similarly, another exact solution to the MKdV Eq. (\ref{8}) is
\be\label{11}
u = A \sqrt{m} \cn[\beta(x-vt+\delta_1),m]\,,
\ee
provided
\be\label{12}
gA^2 = 6\beta^2\,,~~v=(2m-1)\beta^2\,.
\ee

Remarkably, even a linear superposition of the two, i.e.
\be\label{13}
u = \frac{A}{2} \dn[\beta(x-vt+\delta_1),m]
+ \frac{B}{2} \sqrt{m} \cn[\beta(x-vt+\delta_1),m]\,,
\ee
is an exact solution to the MKdV Eq. (\ref{8}) provided
\be\label{14}
B= \pm A\,,~~gA^2 = 6 \beta^2\,,~~v=(1/2)(1+m)\beta^2\,.
\ee
Note that the velocity of the solutions $\dn$, $\cn$ and 
$\dn \pm \sqrt{m} \cn$ is different except at $m=1$.

\subsection{$\phi^2$-$\phi^4$ Model}

We now show that the $\phi^2$-$\phi^4$ field  equation \cite{rajaraman}
\be\label{15}
\phi_{xx}= a \phi +b \phi^3\,,
\ee
also admits such  superposed solutions. 

It is well known that one of the exact solution to the 
field Eq. (\ref{15}) is
\be\label{16}
\phi = A \dn[\beta(x-vt+\delta_1),m]\,,
\ee
provided
\be\label{17}
bA^2 = -2\beta^2\,,~~a=(2-m) \beta^2\,.
\ee
This implies that $a>0,b<0$. 

Another exact solution to the field Eq. (\ref{15}) is
\be\label{18}
\phi = A \sqrt{m} \cn[\beta(x-vt+\delta_1),m]\,,
\ee
provided
\be\label{19}
bA^2 = -2\beta^2\,,~~a=(2m-1)\beta^2\,.
\ee
Thus this solution is valid if $b<0$ while $a > (<)$  0 depending on whether
$m > (<)$ 1/2. Remarkably, even a linear superposition of the two, i.e.
\be\label{20}
\phi = \frac{A}{2} \dn[\beta(x-vt+\delta_1),m]
+ \frac{B}{2} \sqrt{m} \cn[\beta(x-vt+\delta_1),m]\,,
\ee
is an exact solution to the field Eq. (\ref{15}) provided
\be\label{21}
B = \pm A\,,~~bA^2 =- 2 \beta^2\,,~~a=(1/2)(1+m)\beta^2\,.
\ee
Unlike the $\cn$ (but like the $\dn$ solution), these solutions exist only if
$a>0$. 
Note that the value of the width parameter $\beta$ is different for the
three solutions. 

\subsection{$\phi^2$-$\phi^3$-$\phi^4$ case}

The asymmetric double well potential arises in field theory \cite{bazeia} as well 
as in the context of certain first order phase transitions \cite{sanati}. The $\phi^2$- 
$\phi^3$-$\phi^4$ field  equation is \cite{bazeia, sanati} 
\be\label{22}
\phi_{xx}= a \phi +b \phi^2 +c \phi^3\,.
\ee
It admits the periodic solution 
\be\label{23}
u = A +B \dn[\beta(x-vt+\delta_1),m]\,,
\ee
provided
\be\label{24}
A = -\frac{b}{3c}\,,~~cB^2 = -2\beta^2\,,~~a= -2(2-m) \beta^2\,,
~~2b^2 = 9|a||c|\,.
\ee
This implies that $a<0,b>0,c<0$. 

Another exact solution to the field Eq. (\ref{22}) is
\be\label{25}
u = A +B \sqrt{m} \cn[\beta(x-vt+\delta_1),m]\,,
\ee
provided
\be\label{26}
A = -\frac{b}{3c}\,,~~cB^2 = -2\beta^2\,,~~a= -2(2m-1) \beta^2\,,
~~2b^2 = 9|a||c|\,.
\ee
Thus this solution exists only if $b>0, c < 0$ while $a < (>0)$ or = 0 
depending on whether $m > (<)$1/2 or = 0.  

Remarkably, even a linear superposition of the two, i.e.
\be\label{27}
u = A +\frac{B}{2}\dn[\beta(x-vt+\delta_1),m]
+ \frac{D}{2} \sqrt{m} \cn[\beta(x-vt+\delta_1),m]\,,
\ee
is an exact solution to the field Eq. (\ref{22}) provided
\be\label{28}
D = \pm B\,,~~A = -\frac{b}{3c}\,,~~cB^2 = -2 \beta^2\,,~~a= -(m+1) \beta^2\,,
~~2b^2 = 9|a||c|\,.
\ee
Note that as in the $\dn$ case, such solutions exist only if $a,c<0$ 
while $b>0$.

\subsection{Mixed KdV-MKdV system}

The field equations of the mixed KdV-MKdV system are given by
\be\label{29}
u_t+\delta u_{xxx}+\alpha u^2 u_{x} +\gamma u u_{x} =0\,.
\ee
It is easy to show that one of the exact periodic 
solution to Eq. (\ref{29}) is
\be\label{30}
u(x,t) = A+B\dn[\beta(x-vt+\delta_1),m]\,,
\ee
provided 
\be\label{31}
A=-\frac{\gamma}{2\alpha}\,,~~B^2 = \frac{6\delta \beta^2}{\alpha}\,,
\ee
\be\label{32}
v=(2-m)\delta \beta^2 -\frac{\gamma^2}{4\alpha}\,.
\ee

Another periodic solution to Eq. (\ref{29}) is
\be\label{33}
u(x,t) = A+B\sqrt{m} \cn[\beta(x-vt+\delta_1),m]\,,
\ee
provided relations (\ref{31}) are satisfied while the velocity $v$ is given by
\be\label{34}
v=(2m-1)\delta \beta^2 -\frac{\gamma^2}{4\alpha}\,.
\ee

Remarkably, even a superposition of the two solutions (\ref{30}) and (\ref{33})
is also an exact solution but with velocity $v$ which is different than 
that given by either Eq. (\ref{32}) or (\ref{34}). In particular, it is easy 
to show that
\be\label{35}
u(x,t) = A+\frac{B}{2}\dn[\beta(x-vt+\delta_1),m]
+\frac{D}{2} \sqrt{m} \cn[\beta(x-vt+\delta),m]\,,
\ee
is an exact solution to the field Eq. (\ref{29}) 
provided relations (\ref{31}) are satisfied and further
\be\label{36}
D = \pm B\,,~~v=\frac{(1+m)}{2} \delta \beta^2 -\frac{\gamma^2}{4\alpha}\,.
\ee

Notice that the velocities of the three solutions $\dn, \cn$ and 
$\dn \pm \sqrt{m} \cn$ are different.

\subsection{Mixed Quadratic-Cubic NLS Equation}

Let us consider a mixed quadratic-cubic NLS equation \cite{qcnls1, qcnls2} given by 
\be\label{1d}
iu_t+u_{xx}+g_1 |u| u+ g_2|u|^2 u  =0\,.
\ee
One of the exact moving periodic solution to this equation is
\be\label{2d}
u = \big (A \dn[\beta(x-vt+\delta_1),m]+B \big ) 
\exp[-i(\omega t-kx+\delta)]\,,
\ee
provided
\be\label{3d}
g_1 = -3B g_2\,,~~g_2 A^2 = 2\beta^2\,,~~g_2 B^2 = (2-m) \beta^2\,,
~~\omega = k^2 +2(2-m) \beta^2\,,~~v=2k\,.
\ee

Similarly, another exact solution to Eq. (\ref{1d}) is
\be\label{4d}
u = \big (A \sqrt{m} \cn[\beta(x-vt+\delta_1),m]+B \big ) 
\exp[-i(\omega t-kx+\delta)]
\ee
provided
\be\label{5d}
g_1 = -3Bg_2\,,~~g_2 A^2 = 2\beta^2\,,~~g_2B^2 = (2m-1) \beta^2\,,
~~\omega = k^2 +2(2m-1)\beta^2\,,~~v=2k\,.
\ee
Note that this solution exists only if $m > 1/2$.

Remarkably, a linear superposition of the two, i.e.
\be\label{6d}
u = \bigg ( \frac{A}{2} \dn[\beta(x-vt+\delta_1),m]
+ \frac{D}{2} \sqrt{m} \cn[\beta(x-vt+\delta_1),m]+B \bigg )
\exp[-i(\omega t-kx+\delta)]\,,
\ee
is an exact solution to the field  Eq. (\ref{1d}) provided
\bea\label{7d}
&&B = \pm A\,,~~g_1=-3Bg_2\,,~~g_2 A^2 = 2 \beta^2\,, \nonumber \\
&&g_2 B^2 = \frac{1+m}{2} \beta^2\,,~~\omega= k^2 +(1/2)(1+m)\beta^2\,,
~~v=2k\,.
\eea
Note that even though the $\cn$ solution is only valid if $m >1/2$, the 
superposed solution of $\cn$ and $\dn$ is in fact valid over the entire range
of $m$ values, i.e. $0 < m \le 1$. Further, the frequency $\omega$ of the 
three solutions 
(i.e. $\cn, \dn$, and ($\dn \pm \sqrt{m} \cn$) is different except at $m=1$.

\section{Coupled Continuum Field Theories with $\dn \pm \sqrt{m} \cn$
Solutions in Both Fields}
 
We now show that several coupled field theory models (which admit solutions
in terms of $\cn$ and $\dn$ in both the fields) also admit 
$\dn \pm \sqrt{m} \cn$ as solutions in both the fields.
As an illustration, we discuss three examples of coupled continuum field 
theories which admit such solutions.
 
\subsection{Coupled $\phi^4$ Field Theories}

Some time ago, we had considered the coupled $\phi^4$ field
theories with field equations given by \cite{ak1}
\bea\label{37}
&&\phi_{xx} = 2\alpha_1 \phi+4\beta_1 \phi^3+2\gamma \phi \psi^2\,, 
\nonumber \\
&&\psi_{xx} = 2\alpha_2 \psi+4\beta_2 \psi^3+2\gamma \psi \phi^2\,.
\eea
The four well known periodic solutions to these coupled  equations are
$\phi= A\dn$ or $\phi = A\sqrt{m} \cn$ and $\psi= B\dn$ or 
$\psi = B \sqrt{m} \cn$. For illustration, we just discuss one of the known
solutions. The details of the other three solutions can be found in \cite{ak1}. 
For example, one of the solutions is given by 
\be\label{38}
\phi = A\dn[\beta(x+\delta_1),m]\,,~~\psi 
= D\sqrt{m}\cn[\beta(x+\delta_1),m]\,,
\ee
provided
\be\label{39}
\beta^2 = -2\beta_1 A^2 -\gamma D^2 = -2\beta_2 D^2 -\gamma A^2 , 
\ee
\be\label{40}
\alpha_1 = \frac{(2-m)\beta^2 A^2}{2} +\gamma (1-m) D^2\,,~~
\alpha_2 = \frac{(2m-1)\beta^2 D^2}{2} -\gamma (1-m) A^2\,.
\ee
On solving Eq. (\ref{39}) we have
\be\label{39a}
A^2 = \frac{2|\beta_2|-|\gamma|}{4\beta_1 \beta_2 -\gamma^2}\,, ~~~
D^2 = \frac{2|\beta_1|-|\gamma|}{4\beta_1 \beta_2 -\gamma^2}\,.
\ee
In the special case when $\gamma=2\beta_1=2\beta_2 < 0$, instead of the 
relations (\ref{39a}), $A,D$ only satisfy the constraint
\be\label{39b}
\beta^2 = |\gamma| (A^2+D^2)\,.
\ee 

We now show that even a linear superposition of $\dn$ and $\cn$  (in both 
the fields) is an 
exact solution of Eqs. (\ref{37}). In particular, it is easily checked that
\be\label{41}
\phi = \frac{A}{2}\dn[\beta(x+\delta_1),m]
+\frac{B}{2}\sqrt{m}\cn[\beta(x+\delta_1),m]\,,
\ee
\be\label{42}
\psi = \frac{D}{2}\dn[\beta(x+\delta_1),m]
+\frac{E}{2}\sqrt{m}\cn[\beta(x+\delta_1),m]\,,
\ee
is an exact solution to the coupled Eqs. (\ref{37}) provided
\be\label{43}
B=\pm A\,,~~E = \pm D\,, ~~\alpha_1 = \alpha_2 = \frac{(1+m)\beta^2}{4}\,,
\ee
while $A, D$ satisfy Eq. (\ref{39}) and hence relations (\ref{39a}) 
or constraint (\ref{39b}). Note that the signs of $D= \pm A$ and $E = \pm B$
are correlated. 

\subsection{Coupled NLS-MKdV Model}

We now consider a coupled NLS-MKdV system with 
the field equations given by 
\bea\label{a1}
&&iu_t+u_{xx}+g|u|^2 u+ \alpha u v^2=0 \, , \nonumber \\
&&v_t+v_{xxx} +6v^2 v_{x}+\gamma v(|u|^2)_{x} =0\,, 
\eea
where $u$ and $v$ are the NLS and the MKdV fields, respectively. 
As we remarked in Ref. \cite{ak2}, these coupled equations admit four periodic 
solutions with $u$ being either
$\cn$ or $\dn$ (multiplied by an exponential) and similarly $v$ can be
either $\cn$ or $\dn$. However, we only discuss one of the 
four solutions here, given by
\bea\label{a2}
&&u(x,t)=A\exp[-i(\omega t-kx+\delta)] \dn[\beta(x-ct+\delta_1),m]\,,
\nonumber \\
&&v(x,t)=B \sqrt{m} \cn[\beta(x-ct+\delta_1),m]\,,
\eea
provided
\be\label{a3}
c=2k=(2m-1)\beta^2\,,~~\omega=k^2-(2-m)\beta^2+(1-m)\alpha B^2\,,
\ee
\be\label{ax}
gA^2+\alpha B^2 = 2\beta^2\,,~~\gamma A^2+3B^2 = 3\beta^2\,.
\ee
On solving Eqs. (\ref{ax}), we obtain
\be\label{a4}
A^2 = \frac{3(\alpha-2)\beta^2}{\alpha \gamma -3g}\,, ~~~
B^2=\frac{(2\gamma-3g)\beta^2}{(\alpha \gamma -3g)}\,.
\ee
In the special case when $\gamma=(3/2)g,\alpha=2$, $A,B$
remain undetermined, and instead of Eqs. (\ref{a4}) $A,B$
only satisfy the constraint
\be\label{a6}
\gamma A^2+3B^2=3\beta^2\,.
\ee

Remarkably, even a linear superposition
\bea\label{a7}
&&u(x,t)=\frac{1}{2}\exp[-i(\omega t-kx+\delta)] 
\bigg (A\dn[\gamma(x-ct+\delta_1),m] \nonumber \\
&&+D\sqrt{m}\cn[\gamma(x-ct+\delta_1),m] 
\bigg )\,,
\eea
and
\be\label{a8}
v(x,t)= \frac{1}{2} \bigg (B\dn[\gamma(x-ct+\delta_1),m] 
+F\sqrt{m}\cn[\gamma(x-ct+\delta_1),m] \bigg )\,,
\ee
is an exact solution of Eqs. (\ref{a1}) provided 
\be\label{a9}
c=2k=(1+m)\gamma^2/2\,,~~\omega=k(k-2)\,,~~D=\pm A\,,~~F=\pm B\,,
\ee
while $A,B$ are either given by Eqs. (\ref{a4}) or are 
related by the constraint (\ref{a6}). Note that the signs of $D= \pm A$
and $F = \pm B$ are correlated. 

\subsection{Coupled NLS Model}

Let us consider the following coupled NLS field equations
\bea\label{12.1}
&&iu_t+u_{xx}+(a|u|^2+b|v|^2) u =0 \, , \nonumber \\
&&v_t+v_{xx} +(f|u|^2+e|v|^2)v =0\,, 
\eea
where $u$ and $v$ are the two coupled NLS fields. 
Note that in the special case
when $a=f=b=e$ this system reduces to the Manakov system which is a well known
integrable system \cite{man}. Remarkably, even when $a=f=-b=-e$, this is an integrable 
system \cite{mik,zak,ger} which we shall call as MZS (Mikhailov-Zakharov-Schulman) 
system. We shall however discuss the exact periodic solutions of this coupled 
system when the coefficients $a,b,f,e$ are arbitrary but real. 

The coupled equations (\ref{12.1}) admit four 
solutions with either $\cn$ or $dn$ in $u$ as well as $v$ fields and 
several other solutions in terms of Lam\'e polynomials of order 1 and 2. Here, 
as an illustration, we only discuss one such solution and then show that these 
coupled equations also admit solutions in terms of a linear superposition of 
$\dn$ and $\cn$ in both the fields. 
 
It is easily checked that 
\be\label{12.2}
u(x,t)=A \exp[-i(\omega_1 t-k_1 x+\delta_1)] \dn[\beta(x-ct+\delta),m]\,,
\ee
and
\be\label{12.3}
v(x,t)=B \sqrt{m} \exp[-i(\omega_2 t-k_2 x+\delta_2)] \cn[\beta(x-ct+\delta),m]\,,
\ee
is an exact solution to the coupled field equations (\ref{12.1}) provided
\be\label{12.4}
aA^2+bB^2= 2\beta^2\,,~~fA^2+eB^2= 2\beta^2\,,
\ee
and further
\be\label{12.5}
k_1=k_2\,,~~c=2k_1\,,~~\omega_1 = k_1^2 - (2-m)\beta^2-(1-m) bB^2\,,~~
\omega_2 = k_1^2 -(2m-1) \beta^2 -(1-m) fA^2\,.
\ee
On solving Eqs. (\ref{12.4}) we find that so long as $bf \ne ae$, $A,B$ are
given by
\be\label{12.6}
A^2 = \frac{2\beta^2(b-e)}{bf-ae}\,,~~~B^2 = \frac{2\beta^2(f-a)}{bf-ae}\,.
\ee

Few remarks are in order at this stage.
\begin{enumerate}

\item In case $ae=bf$, then along with Eqs. (\ref{12.4}) this implies 
that $b=e$ and $a=f$. In that case instead of the relations (\ref{12.6}), we only
have the constraint 
\be\label{12.7}
aA^2 +b B^2 = 2\beta^2\,. 
\ee

\item In the Manakov case, $a=b=e=f$ and the constraint (\ref{12.7}) becomes 
$a(A^2+B^2) = 2\beta^2$. On the other hand in the MZS case when $a=f=-e=-b$, 
the constraint becomes $a(A^2-B^2)=2\beta^2$. 

\end{enumerate}

Remarkably, it turns out that even a linear superposition of $\dn$ and $\cn$
is a solution to the coupled Eqs. (\ref{12.1}). In particular, 
\be\label{12.8}
u(x,t)= \frac{1}{2} \exp[-i(\omega_1 t-k_1 x+\delta_1)] 
\bigg ( A\dn[\beta(x-ct+\delta),m]
+\sqrt{m} D \cn[\beta(x-ct+\delta),m] \bigg )\,,
\ee
and
\be\label{12.9}
v(x,t)= \frac{1}{2} \exp[-i(\omega_2 t-k_2 x+\delta_2)] 
\bigg ( B \dn[\beta(x-ct+\delta),m]
+\sqrt{m} E \cn[\beta(x-ct+\delta),m] \bigg )\,,
\ee
is an exact solution to the coupled field equations (\ref{12.1}) provided
Eqs. (\ref{12.4}) are satisfied and further
\be\label{12.10}
k_1=k_2\,,~~c=2k_1\,,~~D = \pm A\,,~~E = \pm B\,,~~
\omega_1 = \omega_2 = k_1^2 - \frac{1}{2} (1+m)\beta^2\,.
\ee
Note that the signs of $D = \pm A$ and $E = \pm B$ are correlated.
Further, all the remarks made after the previous solution are also valid for 
this case.

\section{Discrete Nonlinear Equations}

We now discuss five examples of discrete nonlinear equations all of which
are known to admit $\dn$ and $\cn$ as periodic solutions. We show that all 
these models also admit periodic solutions of the form $\dn \pm \sqrt{m} \cn$.

\subsection{Ablowitz-Ladik Model}

It is well known that the celebrated Ablowitz-Ladik (AL) model \cite{al, takeno}, 
which is an integrable model, admits moving
$\dn$ and $\cn$ periodic solutions \cite{bishop}. We now show that the same 
model also admits linearly superposed moving periodic solutions.

We start from the AL equation
\be\label{1.1}
i\frac{du_n}{dt} + u_{n+1}+u_{n-1}+ | u_n|^2 [u_{n+1} +u_{n-1}] = 0\,.
\ee
An exact moving periodic solution to Eq. (\ref{1.1}) is known 
to be 
\be\label{1.2}
u_n = A \dn[\beta(n-vt+\delta_1),m] e^{-i(\omega t -kn +\delta)}\,,
\ee
provided 
\be\label{1.3}
\omega = - \frac{2\cos(k)\dn(\beta,m)}{\cn^2(\beta,m)}\,,~~
\beta v = \frac{2\sin(k)}{\cs(\beta,m)}\,,~~gA^2 \cs^2(\beta,m) =1\,.
\ee

Another exact moving soliton solution to Eq. (\ref{1.1}) is 
\be\label{1.4}
u_n = A\sqrt{m} \cn[\beta(n-vt+\delta_1),m] e^{-i(\omega t -kn +\delta)}\,,
\ee
provided 
\be\label{1.5}
\omega = - \frac{2\cos(k)\cn(\beta,m)}{\dn^2(\beta,m)}\,,~~
\beta v = \frac{2\sin(k)}{\ds(\beta,m)}\,,~~gA^2 \ds^2(\beta,m) =1\,.
\ee

Remarkably, even a linear superposition of the $\dn$ and $\cn$ solutions
 is also an exact
solution to Eq. (\ref{1.1}). In particular, it is easy to show that
\be\label{1.6}
u_n = \bigg (\frac{A}{2}\dn[\beta(n-vt+\delta_1),m]+
\frac{B}{2}\sqrt{m} \cn[\beta(n-vt+\delta_1),m] \bigg ) 
e^{-i(\omega t -kn +\delta)}\,,
\ee
is also an exact solution to Eq. (\ref{1.1}) provided 
\bea\label{1.7}
&&B = \pm A\,,~~\omega = - \frac{4\cos(k)}{(\cn(\beta,m)+\dn(\beta,m)]}\,,
\nonumber \\
&&\beta v = \frac{4\sin(k)}{[\cs(\beta,m)+\ds(\beta,m)]}\,,
~~gA^2 [\cs(\beta,m)+\ds(\beta,m)]^2 =4\,.
\eea
As remarked earlier (in Sec. II), observe that if we replace $\dn(\beta,m)$ and 
$\cn(\beta,m)$ by $[\dn(\beta,m)+\cn\beta,m)]/2$ in relations (\ref{1.3})
and (\ref{1.5}), we recover relations (\ref{1.7}). Exactly the same observation
is also valid in the case of the next four discrete solutions that we
discuss below.

\subsection{Saturable DNLS Equation}

Let us consider the saturable discrete nonlinear Schr\"odinger (DNLS)  equation
\be\label{1.16}
i\frac{du_n}{dt} +[u_{n+1}+u_{n-1}-2u_n]+ \frac{\nu \mid u_n \mid^2 u_n}
{1+ |u_n|^2}u_n = 0\,. 
\ee
It is worth reminding that this equation has received great attention in the 
context of optical pulse propagation in various doped fibers \cite{of}.  It may 
also be relevant for the description of arrays of optical waveguides with 
nonpolynomial nonlinearities and Bose-Einstein condensates \cite{bec}.    

There are two well known periodic solutions to this equation \cite{ak4}.
The first one is
\be\label{1.17}
u_n = A \dn[\beta(n+\delta_1),m] e^{-i(\omega t +\delta)}\,,
\ee
provided
\be\label{1.18}
A^2 \cs^2(\beta,m) =1\,,~~
\omega = 2-\nu = 2\left[1-\frac{\cn(\beta,m)}{\dn^2(\beta,m)}\right]\,.
\ee

The other solution is
\be\label{1.19}
u_n = A\sqrt{m} \cn[\beta(n+\delta_1),m] e^{-i(\omega t +\delta)}\,,
\ee
provided
\be\label{1.20}
A^2 \ds^2(\beta,m) =1\,,~~
\omega = 2 -\nu = 2\left[1-\frac{\dn(\beta,m)}{\cn^2(\beta,m)}\right]\, , 
\ee
where $\cs(\beta,m)=\cn(\beta,m)/\sn(\beta,m)$ and $\ds(\beta,m)=\dn(\beta,m)/\sn(\beta,m)$. 

Remarkably, even a linear superposition of the two is also an exact periodic 
solution to the saturable DNLS Eq. (\ref{1.16}), i.e. it is easy to show that
\be\label{1.21}
u_n = \bigg (\frac{A}{2} \dn[\beta(n+\delta_1),m]+
\frac{B}{2}\sqrt{m} \cn[\beta(n+\delta_1),m] \bigg ) 
e^{-i(\omega t +\delta)}\,,
\ee
is also an exact solution to Eq. (\ref{1.16}) provided
\be\label{1.22}
B = \pm A\,,~~A^2 [\cs(\beta,m)+ \ds(\beta,m)]^2 =4\,,~~
\omega =2 -\nu = 2\left[1-\frac{2}{\dn(\beta,m)+\cn(\beta,m)}\right]\,.
\ee

\subsection{Discrete $\lambda \phi^4$}

Consider the discrete $\lambda \phi^4$  field equation 
\be\label{1.27}
\frac{1}{h^2}[\phi_{n+1}+\phi_{n-1}-2\phi_n]+\lambda \phi_n 
-\frac{\lambda}{2} \phi_n^2[\phi_{n+1}+\phi_{n-1}] =0\, , 
\ee
which is quite similar to the stationary version of the Ablowitz-Ladik equation, 
Eq. (72), except that the field $\phi_n$  is real in the present case.  It is well known 
that Eq. (\ref{1.27})  admits the periodic solution \cite{ak6}
\be\label{1.28}
\phi_n = A \dn[\beta(n+\delta_1),m]\,,
\ee
provided 
\be\label{1.29}
\frac{1}{h^2} = -\frac{\lambda A^2}{2} \cs^2(\beta,m)\,,~~
\lambda - \frac{2}{h^2} = \lambda A^2 \ds(\beta,m) \ns(\beta,m)\,.
\ee
\be\label{1.30}
\Lambda = \lambda h^2 < 0\,,~~A^2 = \frac{2}{|\Lambda|\cs^2(\beta,m)}\,,~~
\Lambda = 2\left[1-\frac{\dn(\beta,m)}{\cn^2(\beta,m)}\right]\,.
\ee

Another known periodic solution to Eq. (\ref{1.27}) is
\be\label{1.31}
\phi_n = A \sqrt{m} \cn[\beta(n+\delta_1),m]\,,
\ee
provided 
\be\label{1.32}
\frac{1}{h^2} = -\frac{\lambda A^2}{2} \ds^2(\beta,m)\,,~~
\lambda - \frac{2}{h^2} = \lambda A^2 \cs(\beta,m) \ns(\beta,m)\,.
\ee
\be\label{1.33}
\Lambda < 0\,,~~A^2 = \frac{2}{|\Lambda|\ds^2(\beta,m)}\,,~~
\Lambda = 2\left[1-\frac{\cn(\beta,m)}{\dn^2(\beta,m)}\right]\,.
\ee

We now show that the same model (\ref{1.27}) also admits superposed solution
of $\cn$ and $\dn$. In particular, it is easy to check that the model admits
an exact solution  
\be\label{1.34}
\phi_n = \frac{A}{2}\dn[\beta(n+\delta_1),m]
+\frac{B}{2} \sqrt{m} \cn[\beta(n+\delta_1),m]\,,
\ee
provided 
\be\label{1.35}
B = \pm A\,,~~\Lambda < 0\,,
~~A^2 = \frac{8}{|\Lambda|[\cs(\beta,m)+\ds(\beta,m)]^2}\,,~~
\Lambda = 2\left[1-\frac{2}{\cn(\beta,m) \dn(\beta,m)}\right]\,.
\ee

\subsection{Discrete Cubic-Quintic Model}

There is a relation between the continuum generalized NLS and the cubic-quintic NLS 
\cite{kody}.  In the present discrete case, the discrete field equations are 
\be\label{1.36}
i\frac{du_n}{dt} + [u_{n+1}+u_{n-1}]+ g_1 | u_n|^4 [u_{n+1} +u_{n-1}] 
+g_2 |u_n|^2 u_n = 0\,.
\ee
An exact periodic solution to Eq. (\ref{1.36}) is known to be 
\cite{ak7} 
\be\label{1.37}
u_n = A \dn[\beta(n+\delta_1),m] e^{-i(\omega t +\delta)}\,,
\ee
provided 
\be\label{1.38}
g_1 <0\,,~~\omega = - 2\frac{\dn(\beta,m)}{\cn^2(\beta,m)}\,,
~~|g_1| A^4 \cs^4(\beta,m) =1\,,
~~\frac{g_2}{2\sqrt{|g_1|}} = \frac{\dn(\beta,m)}{\cn^2(\beta,m)}\,.
\ee

Another exact periodic solution to Eq. (\ref{1.36}) is
\be\label{1.39}
u_n = A\sqrt{m} \cn[\beta(n+\delta_1),m] e^{-i(\omega t +\delta)}\,,
\ee
provided 
\be\label{1.40}
g_1 <0\,,~~\omega = - 2\frac{\cn(\beta,m)}{\dn^2(\beta,m)}\,,
~~|g_1| A^4 \ds^4(\beta,m) =1\,,
~~\frac{g_2}{2\sqrt{|g_1|}} = \frac{\cn(\beta,m)}{\dn^2(\beta,m)}\,.
\ee

Remarkably, even a linear superposition of the above two is also an exact
solution to the Eq. (\ref{1.36}). In particular, it is easy to show that
\be\label{1.41}
u_n = \bigg (\frac{A}{2}\dn[\beta(n+\delta_1),m]+
\frac{B}{2}\sqrt{m} \cn[\beta(n+\delta_1),m] \bigg ) 
e^{-i(\omega t +\delta)}\,,
\ee
is also a periodic solution to Eq. (\ref{1.36}) provided 
\bea\label{1.42}
&&g_1<0\,,~~\omega = - \frac{4}{(\cn(\beta,m)+\dn(\beta,m)]}\,,~~
~~|g_1| A^4 [\ds(\beta,m)+\cs(\beta,m)]^4 =16\,, \nonumber \\
&&B = \pm A\,,~~\frac{g_2}{2\sqrt{|g_1|}} = \frac{2}{\dn(\beta,m)+\cn(\beta,m)}\,.
\eea

\subsection{Discrete MKdV Model}

The discrete MKdV equation is known to be an integrable equation \cite{al} 
 and is given by 
\be\label{1.1a}
\frac{du_n}{dt} +\alpha[u_{n+1}+u_{n-1}]+  u_n^2 [u_{n+1} - u_{n-1}] = 0\,,
\ee
where $u_n(t)$ is a real field. We now show that not only the real but even
the complex discrete MKdV equation  
\be\label{1.1b}
\frac{du_n}{dt} +\alpha[u_{n+1}+u_{n-1}]+  |u_n|^2 [u_{n+1} - u_{n-1}] = 0\,,
\ee
has such superposed solutions.

An exact moving periodic solution to Eq. (\ref{1.1b}) is  
\be\label{1.2a}
u_n = A \dn[\beta(n-vt+\delta_1),m] e^{-i(\omega t -kn +\delta)}\,,
\ee
provided 
\be\label{1.3a}
\omega =   \frac{2\alpha \sin(k)\dn(\beta,m)}{\cn^2(\beta,m)}\,,~~
\beta v = \frac{2\alpha \cos(k)}{\cs(\beta,m)}\,,
~~A^2 \cs^2(\beta,m) = \alpha > 0\,.
\ee
It may be noted that in the limit $k=0$, $\omega=0$, the solution (\ref{1.2a}) 
is an exact solution to the real, discrete MKdV Eq. (\ref{1.1a}). 

Another exact moving soliton solution to Eq. (\ref{1.1b}) is 
\be\label{1.4a}
u_n = A\sqrt{m} \cn[\beta(n-vt+\delta_1),m] e^{-i(\omega t -kn +\delta)}\,,
\ee
provided 
\be\label{1.5a}
\omega =   \frac{2\alpha \sin(k)\cn(\beta,m)}{\dn^2(\beta,m)}\,,~~
\beta v = \frac{2\alpha \cos(k)}{\ds(\beta,m)}\,,
~~A^2 \ds^2(\beta,m) = \alpha > 0\,.
\ee
In the limit $k=\omega =0$, it reduces to an exact solution of the real, 
discrete MKdV Eq. (\ref{1.1a}). 

Remarkably, even a linear superposition of the $\dn$ and $\cn$ solutions
 is also an exact
solution to Eq. (\ref{1.1b}). In particular, it is easy to show that
\be\label{1.6a}
u_n = \bigg (\frac{A}{2}\dn[\beta(n-vt+\delta_1),m]+
\frac{B}{2}\sqrt{m} \cn[\beta(n-vt+\delta_1),m] \bigg ) 
e^{-i(\omega t -kn +\delta)}\,,
\ee
is also an exact  solution to Eq. (\ref{1.1b}) provided 
\bea\label{1.7a}
&&B = \pm A\,,~~\omega = \frac{4\alpha \sin(k)}{[(\cn(\beta,m)+\dn(\beta,m)]}\,,
\nonumber \\
&&\beta v = \frac{4 \alpha \cos(k)}{[\cs(\beta,m)+\ds(\beta,m)]}\,,
~~A^2 [\cs(\beta,m)+\ds(\beta,m)]^2 =4 \alpha > 0\,.
\eea
In the limit $k=\omega =0$, it reduces to an exact solution of the real, 
discrete MKdV Eq. (\ref{1.1a}). 

While deriving the various solutions in this section, several not so well
known identities for the Jacobi Elliptic Functions (JEF) have been used 
which have been obtained by one us a few years ago \cite{ak3}; 
they are given in the Appendix.

\section{$\dn^2 \pm \sqrt{m} \cn \dn$ superposed Solutions}

We now discuss three examples of continuum field theories which admit $\dn^2$
as a solution and show that all three models also admit solutions of the form 
$\dn^2\pm \sqrt{m} \cn \dn$, even though $\cn \dn$ is not a solution of any of
these models. 

\subsection{KdV Equation}

We first discuss the celebrated KdV equation
\be\label{2.1}
u_t+u_{xxx}+gu u_{x} =0\,,
\ee
which is a well known integrable equation having application in several areas
including shallow water waves \cite{1,2}. It is well known that it admits periodic 
soliton solution of the form $\dn^2$. We now show that it also admits  superposed 
solutions of the form $dn^2 \pm \sqrt{m} \cn \dn$ even though $\cn \dn$ is 
not a solution of the KdV equation. 

It is well known that one of the exact solution to the 
KdV Eq. (\ref{2.1}) is
\be\label{2.2}
u = A \dn^2[\beta(x-vt+\delta_1),m]\,,
\ee
provided
\be\label{2.3}
gA = 12 \beta^2\,,~~v= 4(2-m) \beta^2\,.
\ee

Remarkably, even 
\be\label{2.4}
u = \frac{1}{2} \bigg (A\dn^2[\beta(x-vt+\delta_1),m]+
B \sqrt{m} \cn[\beta(x-vt+\delta_1),m] \dn[\beta(x-vt+\delta_1),m] \bigg )\,,
\ee
is an exact solution of the KdV Eq. (\ref{2.1}) provided
\be\label{2.5}
B = \pm A\,,~~gA = 12 \beta^2\,,~~v= (5-m)\beta^2\,.
\ee
It is worth repeating once again that $\cn \dn$ is {\it not} an exact solution
to the KdV Eq. (\ref{2.1}). 
We thus have two new periodic solutions of KdV Eq. (\ref{2.1}) 
depending on whether $B=A$ or $B=-A$. 

Several remarks are in order here which are in fact valid for all the 
solutions of the form $\dn^2 \pm \sqrt{m} \cn\dn$ which we discuss in this
and section and Sec. VI.
Therefore, we shall not repeat these remarks while discussing various
solutions in these sections.

\begin{enumerate}

\item All the models discussed in this and the Sec. VI 
admit $\dn^2$ as well as 
$\dn^2 \pm \sqrt{m}\cn \dn$ as solutions even though $\sqrt{m} \cn \dn$
is not a solution of any of these models.

\item In the limit $m=1$, the two solutions $\dn^2$ and  
$\dn^2+\sqrt{m}\cn\dn$ go over to the well known pulse (i.e. $\sech^2$) 
solution while $\dn^2-\sqrt{m} \cn \dn$ 
solution goes over to the vacuum solution. 

\item In all the  cases discussed in this and the next section, the factors of 
$2-m$ and $1-m+m^2$ 
which appear in the $\dn^2$ solution, get
replaced by the factors of $(5-m)/4$ and $\sqrt{1+14m+m^2}/4$, respectively,  
in the $\dn^2 \pm \sqrt{m} \cn \dn$ solutions. 

\end{enumerate}

\subsection{Quadratic NLS Equation}

We show that the quadratic NLS equation given by 
\be\label{2.6}
iu_t+u_{xx}+g|u| u  =0\,,
\ee
not only admits $\dn^2+D$ as a solution but it also admits the superposed 
solution of the form $\dn^2 \pm \sqrt{m} \cn \dn+D$.
 
It is easily checked that one of the exact solution to Eq. (\ref{2.6}) is
\be\label{2.7}
u = \bigg (A \dn^2[\beta(x-vt+\delta_1),m]+D \bigg ) 
\exp[-i(\omega t-kx+\delta)]\,,
\ee
provided
\be\label{2.8}
gA = 6 \beta^2\,,~~\omega= k^2 +4(2-m)\beta^2+2gD\,,~~v=2k\,,
\ee
and 
\be\label{2.9}
gD=-2[(2-m) \pm  \sqrt{1-m+m^2}]\beta^2\,.
\ee

Remarkably, even 
\bea\label{2.10}
&&u = \bigg ( D+\frac{A}{2} \dn^2[\beta(x-vt+\delta_1),m] \nonumber \\
&&+ \frac{B}{2} \sqrt{m} \cn[\beta(x-vt+\delta_1),m] 
\dn[\beta(x-vt+\delta_1),m] \bigg )
\exp[-i(\omega t-kx+\delta)]\,,
\eea
is an exact solution to Eq. (\ref{2.6}) provided
\be\label{2.11}
v=2k\,,~~B = \pm A\,,~~gA = 6 \beta^2\,,~~\omega = k^2 +(5-m)\beta^2
+2gD\,,
\ee
and further
\be\label{2.12}
2gD=-[(5-m) \pm  \sqrt{1+14m+m^2}]\beta^2\,.
\ee
Again note that $\cn \dn$ is not an exact solution to Eq. (\ref{2.6}). 

\subsection{$\phi^3$ Field Theory}

This field theory arises in the context of third order phase transitions \cite{phi3} 
and is also relevant to tachyon condensation \cite{tachyon}. It is well known that 
the field equation for the $\phi^2-\phi^3$ field theory given by
\be\label{2.13}
\phi_{xx} = a\phi+b\phi^2\,,
\ee
admits an exact periodic solution
\be\label{2.14}
\phi = A\dn^2[\beta(x+c),m]+B\,,
\ee
provided
\be\label{2.15}
A=-\frac{3a}{2b\sqrt{1-m+m^2}}\,,~~\beta^2=\frac{a}{4\sqrt{1-m+m^2}}\,,~~
B=\frac{a[2-m-\sqrt{1-m+m^2}]}{2b\sqrt{1-m+m^2}}\,.
\ee

Remarkably, we find that the same model also admits the superposed 
periodic solution
\be\label{2.16}
\phi = \frac{A}{2}\dn^2[\beta(x+c),m]+\frac{D}{2}\sqrt{m} \cn[\beta(x+c),m]
\dn[\beta(x+c),m]+B\,,
\ee
provided
\be\label{2.17}
D=A =-\frac{6a}{b\sqrt{1+14m+m^2}}\,,~~\beta^2=\frac{a}{\sqrt{1+14m+m^2}}\,,~~
B=\frac{a[5-m-\sqrt{1+14m+m^2}]}{2b\sqrt{1+14m+m^2}}\,.
\ee

\section{Coupled Field Theories with $\dn^2 \pm \sqrt{m} \cn\dn$ 
Solutions}

We now discuss three examples of coupled field theories which are known to 
admit $\dn^2$ as a periodic solution in both the fields, and show that these 
coupled models also admit $\dn^2 \pm \sqrt{m} \cn \dn$ as solutions in both 
the fields.

\subsection{Quadratic NLS-KdV Coupled Theory}

We first consider the quadratic NLS-KdV (QNLS-KdV)  coupled system.
The field equations for the coupled QNLS-KdV system are
\bea\label{2.18}
&&iu_t+u_{xx}+g_1 |u| u+\alpha u v  =0\,, \nonumber \\
&&v_t+v_{xxx}+6 vv_{x}+\gamma v|u|_{x}  =0\,,
\eea
where $u$ and $v$ denote the QNLS and KdV fields, respectively. 
It is easily shown that this coupled system admits an exact solution
\bea\label{2.19}
&&u = \bigg (A \dn^2[\beta(x-ct+\delta_1),m]+D \bigg ) 
\exp[-i(\omega t-kx+\delta)]\,,
 \nonumber \\
&&v = B \dn^2[\beta(x-ct+\delta_1),m]+F\,,
\eea
provided
\be\label{2.20}
gA +\alpha B = 6 \beta^2\,,~~\gamma A+6B = 12 \beta^2\,,
\ee
and further
\be\label{2.21}
c=2k\,,~~\omega= k^2-2[2(2-m)+3(z+y)]+Ag(z-y)\,,~~c=4[2-m+3z]\beta^2\,,
\ee
with 
\be\label{2.22}
z=\frac{F}{B}\,,~~y = \frac{D}{A} = 
\frac{-(2-m) \pm \sqrt{m^2-2(1-m)}}{3}\,.
\ee
On solving the relations (\ref{2.20}), we find that in general $A,B$
are given by
\be\label{2.23}
A=\frac{12(3-\alpha)\beta^2}{6g-\alpha \gamma}\,, ~~~
B=\frac{6(2g-\gamma)\beta^2}{6g-\alpha \gamma}\,.
\ee
Only in the special case when $\gamma=2g, \alpha=3$ that $A,B$ cannot be
separately determined but they only satisfy the constraint
\be\label{2.24}
3B+gA=6\beta^2\,.
\ee

Remarkably, even a superposition, i.e.
\bea\label{2.25}
&&u = \bigg ( D+\frac{A}{2} \dn^2[\beta(x-ct+\delta_1),m] \nonumber \\
&&+ \frac{G}{2} \sqrt{m} \cn[\beta(x-ct+\delta_1),m] 
\dn[\beta(x-vt+\delta_1),m] \bigg )
\exp[-i(\omega t-kx+\delta)]\,,
\eea
and
\bea\label{2.25a}
&&v = F+\frac{B}{2} \dn^2[\beta(x-ct+\delta_1),m] \nonumber \\
&&+ \frac{H}{2} \sqrt{m} \cn[\beta(x-ct+\delta_1),m] 
\dn[\beta(x-ct+\delta_1),m]\,,
\eea
is an exact solution to Eqs. (\ref{2.18}) provided relations (\ref{2.20})
(and hence (\ref{2.23}) or (\ref{2.24}) are satisfied) and 
further                  
\bea\label{2.26}
&&c=2k\,,G= \pm A\,,~~H = \pm B\,,~~c=[5-m+12z]\,,~~z=\frac{F}{B}\,, 
\nonumber \\
&&\omega=k^2-[5-m+6(z+y)]\beta^2+gA(z-y)\,,~~
y=\frac{D}{A}=\frac{-(5-m)\pm \sqrt{1+14m+m^2}}{12}\,.
\eea
Note that the signs of $H = \pm B$ and $G = \pm A$ are correlated.

\subsection{NLS-MKdV Coupled Field Theory}

In section III we have discussed coupled
NLS-MKdV system and  shown that it admits $\cn$, $\dn$ as well as 
superposed solutions of the form $\dn \pm \sqrt{m} \cn$
in both the fields. We now show that remarkably, the same coupled model 
not only admits $\dn^2$ as a solution in both the fields but it also admits 
$\dn^2 \pm \sqrt{m}\cn \dn$ as a solution in both the fields  even
though neither $\dn^2$ nor $\cn \dn$ nor $\dn^2 \pm \sqrt{m} \cn \dn$ 
is an exact solution of either of the uncoupled NLS or MKdV models.

The field equations for the coupled NLS-MKdV system as given in Sec. III 
(see Eqs. (\ref{a1}))  are
\bea\label{2.27}
&&iu_t+u_{xx}+g_1 |u|^2 u+\alpha u v^2  =0\,, \nonumber \\
&&v_t+v_{xxx}+6 v^2 v_{x}+\gamma v(|u|^2)_{x}  =0\,,
\eea
where $u$ and $v$ denote the NLS and the MKdV fields, respectively.
It is easily shown that this coupled system admits an exact solution
\bea\label{2.28}
&&u = \bigg (A \dn^2[\beta(x-ct+\delta_1),m]+F \bigg ) 
\exp[-i(\omega t-kx+\delta)]\,,
 \nonumber \\
&&v = B \dn^2[\beta(x-ct+\delta_1),m]+D\,,
\eea
provided
\bea\label{2.29}
&&\alpha =3/2\,,~~\gamma A^2 = -3B^2\,,~~\gamma=2g <0\,,~~(z-y)B^2=2\beta^2\,,
\nonumber \\
&&c=2k=4[(2-m)+3z]\beta^2\,,~~\omega = k^2-[4(2-m)+9y+3z]\beta^2\,, 
\nonumber \\
&&z= \frac{D}{B}\,,~~y=\frac{F}{A}=  \frac{[-(2-m) \pm \sqrt{1-m+m^2}]}{3}\,.
\eea
Note that $A\dn^2+B$ is not a solution of either the NLS or the MKdV uncoupled 
field equations even though the coupled system admits such a solution in both
the fields.

Remarkably, even a superposition, i.e.
\bea\label{2.30}
&&u = \bigg ( F+\frac{A}{2} \dn^2[\beta(x-ct+\delta_1),m] \nonumber \\
&&+ \frac{G}{2} \sqrt{m} \cn[\beta(x-ct+\delta_1),m] 
\dn[\beta(x-vt+\delta_1),m] \bigg )
\exp[-i(\omega t-kx+\delta)]\,,
\eea
and
\bea\label{2.31}
&&v = D+\frac{B}{2} \dn^2[\beta(x-ct+\delta_1),m] \nonumber \\
&&+ \frac{H}{2} \sqrt{m} \cn[\beta(x-ct+\delta_1),m] 
\dn[\beta(x-ct+\delta_1),m]\,,
\eea
is an exact solution to Eq. (\ref{2.27}) provided 
\bea\label{2.32}
&&G = \pm A\,,~~H = \pm B\,,~~c=2k=[5-m+12z]\beta^2\,, \nonumber \\
&&\omega = k^2-[(5-m)+9y+3z]\beta^2\,, 
~~y=\frac{F}{A}=  \frac{[-(5-m) \pm \sqrt{1+14m+m^2}]}{12}\,,
\eea
while rest of the relations are exactly the same as those given by 
Eq. (\ref{2.29}).
Again the signs of $G = \pm A$ and $H = \pm B$ are correlated.

\subsection{Coupled NLS Models}

In section III we have discussed a coupled
NLS system and  shown that it admits $\cn$, $\dn$ as well as 
superposed solutions of the form $\dn \pm \sqrt{m} \cn$
in both the fields. We now show that remarkably, the same 
coupled model also admits 
$\dn^2$ type as well as 
$\dn^2 \pm \sqrt{m}\cn \dn$ type solutions in both the fields even
though neither $\dn^2$ nor $\cn \dn$ nor $\dn^2 \pm \sqrt{m} \cn \dn$ 
is an exact solution of the uncoupled NLS equation.

The field equations for the coupled NLS system as given in Sec. III 
(see Eqs. (\ref{12.1}))  are
\bea\label{2.33}
&&iu_t+u_{xx}+[a|u|^2+b|v|^2] u =0 \, , \nonumber \\
&&v_t+v_{xx} +[f|u|^2+e|v|^2]v =0\,, 
\eea
where $u$ and $v$ are the two coupled NLS fields. 

It is easily checked that 
\be\label{2.34}
u(x,t)=\exp[-i(\omega_1 t-k_1 x+\delta_1)] \big (A \dn^2[\beta(x-ct+\delta),m]
+D \big )\,,
\ee
and
\be\label{2.35}
v(x,t)= \exp[-i(\omega_2 t-k_2 x+\delta_2)] 
\big (B \dn^2[\beta(x-ct+\delta),m]+E \big )\,,
\ee
is an exact solution to the coupled field equations (\ref{2.33}) provided
\be\label{2.36}
c=2k_1=2k_2\,,~~a=f\,,~~b=e\,,~~ a A^2 = - b B^2\,,~~(z_{\pm}-y_{\mp})aA^2 
= 6\beta^2\,,~~z=\frac{D}{A}\,,~~y=\frac{E}{B}\,,
\ee
\bea\label{2.37}
&&z_{\pm}=y_{\pm} =\frac{-(2-m) \pm \sqrt{1-m+m^2}}{3}\,, \nonumber \\
&&\omega_1 -k_1^2 = \mp 2 \sqrt{1-m+m^2} \beta^2\,,~~\omega_2 -k_1^2 =  
\pm 2 \sqrt{1-m+m^2} \beta^2\,.
\eea
From the relation (\ref{2.36}) it follows that this solution exists only if 
$z$ and $y$ are unequal. Also, depending on whether we choose $z_{+},y_{-}$ or 
$z_{-},y_{+}$, the corresponding $\omega_1,\omega_2$ are as given by 
Eq. (\ref{2.37}).

Note that this solution is only valid if $a,b$ have opposite signs, hence this 
solution can only be valid in the MZS case \cite{mik, zak, ger} but not in the 
Manakov case \cite{man}.

Remarkably, it turns out that even a supperposition of $\dn^2$ and $\cn \dn$ 
is an exact solution to the coupled Eqs. (\ref{2.33}). In particular, it is 
easily checked that 
\bea\label{2.38}
&&u(x,t)= \exp[-i(\omega_1 t-k x+\delta_1)] \bigg (\frac{A}{2} 
\dn^2[\beta(x-ct+\delta),m] \nonumber \\
&&+D + \frac{G}{2} \sqrt{m} \cn[\beta(x-ct+\delta),m] 
\dn[\beta(x-ct+\delta),m] \bigg )\,,
\eea
and
\bea\label{2.39}
&&v(x,t)= \exp[-i(\omega_2 t-k x+\delta_1)] \bigg (\frac{B}{2} 
\dn^2[\beta(x-ct+\delta),m] \nonumber \\
&&+E + \frac{H}{2} \sqrt{m} \cn[\beta(x-ct+\delta),m] 
\dn[\beta(x-ct+\delta),m] \bigg )\,,
\eea
is an exact solution to the coupled field equations (\ref{2.33}) provided
\bea\label{2.40}
&&c=2k_1=2k_2\,,~~G= \pm A\,,~~H = \pm B\,,~~a=f\,,~~b=e\,,~~ a A^2 = - b B^2\,,
\nonumber \\
&&(z_{\pm}-y_{\mp})aA^2 
= 6\beta^2\,,~~z=\frac{D}{A}\,,~~y=\frac{E}{B}\,,
\eea
\bea\label{2.41}
&&z_{\pm}=y_{\pm} =\frac{-(5-m) \pm \sqrt{1+14m+m^2}}{12}\,, \nonumber \\
&&\omega_1 -k_1^2 
= \mp \frac{\sqrt{1-m+m^2}}{2} \beta^2\,,~~\omega_2 -k_2^2 =  
\pm \frac{\sqrt{1-m+m^2}}{2} \beta^2\,.
\eea
Thus like the $\dn^2$ solution, this solution exists only if 
$z$ and $y$ are unequal. Also, depending on whether we choose $z_{+},y_{-}$ or 
$z_{-},y_{+}$, the corresponding $\omega_1,\omega_2$ are as given by 
Eq. (\ref{2.41}). Further, like the last solution, this solution is only 
valid if $a,b$ have opposite signs, hence this solution can only be valid 
in the MZS case \cite{mik, zak, ger} but not in the Manakov case \cite{man}.

\section{Mixed $\dn^2\pm \sqrt{m}\cn\dn$ and $\dn \pm \sqrt{m} \cn$ Solutions in 
Coupled Field Theories}

We consider four coupled field theory models in which one field has a solution 
in terms of either $\cn$ or $\dn$ while the other field admits a solution in 
terms of $\dn^2$. We now show that these models also admit superposed 
solutions of the form $\dn\pm \sqrt{m} \cn$ in one field and 
$\dn^2 \pm \sqrt{m} \cn \dn$ in the other field.

\subsection{Coupled NLS-KdV Fields}

Let us consider the following coupled NLS-KdV field equations
\bea\label{6.18}
&&iu_t+u_{xx}+g|u|^2 u+ \alpha u v=0 \, , \nonumber \\
&&v_t+v_{xxx} +6vv_{x}+\gamma v(|u|^2)_{x} =0\,, 
\eea
where $u$ and $v$ are the NLS and the KdV fields, respectively.  It is worth
pointing out that this coupled model has been popular in the literature
in the context of interaction (between a short wave and a long wave) in 
fluid mechanics and plasma physics \cite{bb}.
We first show that these coupled equations admit solutions in terms of $\dn$
and $\cn$ for the NLS field and $\dn^2$ for the KdV field.

It is easily checked that 
\be\label{6.19}
u(x,t)=A\exp[-i(\omega t-kx+\delta_1)] \dn[\beta(x-ct+\delta),m]\,,
\ee
and
\be\label{6.20}
v(x,t)=B \dn^2[\beta(x-ct+\delta_1),m]+D\,,
\ee
is an exact solution to the coupled Eqs. (\ref{6.18}) provided
\be\label{6.20a}
gA^2 +\alpha B = 2 \beta^2\,,~~\gamma A^2 + 6B = 12 \beta^2\,,
\ee
\be\label{6.21}
c=2k=4[2-m+3z]\beta^2\,,~~\omega= k^2-(2-m)\beta^2 -\alpha z B\,, ~~
z=\frac{D}{B} . 
\ee
On solving Eqs. (\ref{6.20a}) we find that 
\be\label{6.22}
A^2 = \frac{12(\alpha-1)\beta^2}{\alpha \gamma -6g}\,,
~~~B=\frac{2(\gamma-6g)\beta^2}{(\alpha \gamma -6g)}\,.
\ee
In the special case when $\alpha = 1, \beta = 6g$,
$A$ and $B$ are undetermined and instead of the relations (\ref{6.22}), 
we only have the constraint
\be\label{6.23}
B+gA^2=2\gamma^2\,.
\ee

It is easily checked that 
\be\label{6.25}
u(x,t)=A\exp[-i(\omega t-kx+\delta)] \sqrt{m} \cn[\gamma(x-ct+\delta_1),m]\,,
\ee
and
\be\label{6.26}
v(x,t)=B   \dn^2[\gamma(x-ct+\delta_1),m]+D\,,
\ee
is an exact solution to the coupled field equations (\ref{6.18}) provided
\be\label{6.27}
\omega=k^2-(2m-1)\gamma^2-[\alpha z+(1-m)]B\,,
\ee
while all other relations are exactly as given by Eqs. (\ref{6.20a}) to
(\ref{6.23}). 

Remarkably, the same model also admits interesting superposed solutions 
of the form $\dn \pm \sqrt{m} \cn$ in the NLS field and $\dn^2 \pm \sqrt{m} \cn\dn$ 
in the KdV field.

It is easily checked that 
\bea\label{6.28}
&&u(x,t)=\frac{1}{2}\exp[-i(\omega t-kx+\delta)] 
\bigg (A\dn[\gamma(x-ct+\delta_1),m]
\nonumber \\
&&+H\sqrt{m}\cn[\gamma(x-ct+\delta_1),m] \bigg )\,,
\eea
and
\bea\label{6.29}
&&v(x,t)=\frac{B}{2} \dn^2[\gamma(x-ct+\delta_1),m] \nonumber \\
&&+\frac{F}{2} \sqrt{m} \cn[\gamma(x-ct+\delta_1),m] 
\dn[\gamma(x-ct+\delta_1),m]+D\,,
\eea
is an exact solution to the coupled field equations (\ref{6.18}) provided
\bea\label{6.30}
&&z=\frac{D}{B}\,,~~
H=\pm A\,,~~F= \pm B\,,~~c=2k=[(5-m)+12z]\gamma^2\,, \nonumber \\
&&\omega= k^2-\frac{(1+m)\gamma^2}{2}-\frac{\alpha}{4}[(1-m)+4z] B\,,
\eea
and further $A,B$ are again either given by Eqs. (\ref{6.22})  
or satisfy the constraint (\ref{6.23}). Note that the signs of $F = \pm B$ and 
$H = \pm A$ are correlated.

\subsection{KdV-MKdV coupled System}

Let us consider the following coupled KdV-MKdV field equations
\bea\label{6.31}
&&u_t+u_{xxx}+6uu_{x}+ 2\alpha u v v_{x} =0 \, , \nonumber \\
&&v_t+v_{xxx} +6v^2 v_{x}+\gamma v u_{x} =0\,, 
\eea
where $u$ and $v$ are the KdV and the MKdV fields, respectively. We first 
show that these coupled equations admit $\dn^2$-$\dn$ and $\dn^2$-$\cn$
solutions.

It is easily checked that 
\be\label{6.32}
u(x,t)=A\dn^2[\beta(x-ct+\delta_1),m]+B\,,
\ee
and
\be\label{6.33}
v(x,t)=D \dn[\beta(x-ct+\delta_1),m]\,,
\ee
is an exact solution to the coupled field equations (\ref{6.31}) provided
\be\label{6.34a}
6A+\alpha D^2 = 12 \beta^2\,,~~\gamma A + 3D^2 = 3 \beta^2\,,
\ee
and further
\be\label{6.35}
c = (2-m)\beta^2\,,~~\frac{B}{A}=-\frac{2-m}{4}\,.
\ee
On solving Eqs. (\ref{6.34a}) we obtain
\be\label{6.34}
 D^2 = \frac{6(2\gamma-3)\beta^2}{\alpha \gamma -18}\,,~~
A = \frac{3(\alpha-12)\beta^2}{\alpha \gamma -18}\,.
\ee
In the special case when $\gamma = 3/2, \alpha = 12$,
$A$ and $D$ are undetermined and instead of the relations (\ref{6.34}), 
we only have the constraint
\be\label{6.35a}
2D^2+A=2\beta^2\,.
\ee

It is easily checked that 
\be\label{6.36}
u(x,t)=A\dn^2[\beta(x-ct+\delta_1),m]+B\,,
\ee
and
\be\label{6.37}
v(x,t)=D \sqrt{m} \cn[\beta(x-ct+\delta_1),m]\,,
\ee
is an exact solution to the coupled field equations (\ref{6.31}) provided
$A,D$ satisfy Eq. (\ref{6.34}) or the constraint (\ref{6.35a}), 
while the other two relations now are
\be\label{6.38}
c = (2m-1)\beta^2\,,~~\frac{B}{A}=-\frac{3-2m}{4}\,.
\ee

We now show that remarkably, these coupled equations admit even a linear 
superposition of the two, i.e.
$\dn^2 \pm \sqrt{m} \cn\dn$ as a solution of the KdV field and $\dn \pm \sqrt{m}\cn$ 
as solution of the MKdV field.  It is easily checked that 
\bea\label{6.39}
&&u(x,t)=\frac{A}{2} \dn^2[\beta(x-ct+\delta_1),m] \nonumber \\
&&+ \frac{F}{2} \sqrt{m} 
\cn[\beta(x-ct+\delta_1),m] \dn[\beta(x-ct+\delta_1),m]+B\,,
\eea
and
\be\label{6.40}
v(x,t)=\frac{D}{2} \dn[\beta(x-ct+\delta_1),m]
+\frac{G}{2} \sqrt{m} \cn[\beta(x-ct+\delta_1),m]\,,
\ee
is an exact solution to the coupled field equations (\ref{6.31}) provided
\be\label{6.41}
F=\pm A\,,~~G=\pm D\,,~~c = \frac{(1+m)}{2}\beta^2\,,
~~\frac{B}{A}=-\frac{3-m}{8}\, , 
\ee
while $A,D$ satisfy the relations (\ref{6.34}) or the constraint (\ref{6.35a}).
Note that the signs of $F = \pm A$ and $G = \pm D$ are correlated.

\subsection{Quadratic NLS-MKdV Coupled Model}

Let us consider a coupled QNLS-MKdV model with the field equations
\bea\label{6.42}
&&iu_t+u_{xx}+g|u|u + \alpha u v^2  =0 \, , \nonumber \\
&&v_t+v_{xxx} +6v^2 v_{x}+\gamma v |u|_{x} =0\,, 
\eea
where $u$ and $v$ are the QNLS and the MKdV fields, respectively.  We first 
show that these coupled equations admit $\dn^2$-$\dn$ and $\dn^2$-$\cn$
solutions.

It is easily checked that 
\be\label{6.43}
u(x,t)= \bigg (A\dn^2[\beta(x-ct+\delta_1),m]+B \bigg )
e^{-i[\omega t-kx+\delta]}\,,
\ee
and
\be\label{6.44}
v(x,t)=D \dn[\beta(x-ct+\delta_1),m]\,,
\ee
is an exact solution to the coupled field equations (\ref{6.42}) provided
\be\label{6.45}
\alpha D^2+gA =6\beta^2\,,~~3D^2+\gamma A = 3\beta^2\,,
\ee
and further
\bea\label{6.46}
&&c = 2k= (2-m)\beta^2\,,~~z = \frac{B}{A}=\frac{-(2-m)\pm \sqrt{1-m+m^2}}{3}
\,, \nonumber \\
&&\omega = k^2-2[2(2-m)+3z]\beta^2-Agz\,.
\eea
On solving Eqs. (181) we find that $A,D^2$ are given by 
\be\label{6.47}
A = \frac{3(6-\alpha)\beta^2}{3g-\alpha \gamma}\,,~~~ 
 D^2 = \frac{3(g-2\gamma)\beta^2}{3g-\alpha \gamma}\,.  
\ee
However, in the special case when $2\gamma = g, \alpha = 6$,
$A$ and $D$ are undetermined and instead of the relations (\ref{6.47}), 
we only have the constraint
\be\label{6.48}
3D^2+\gamma A=3\beta^2\,.
\ee

It is easily checked that 
\be\label{6.49}
u(x,t)= \bigg (A\dn^2[\beta(x-ct+\delta_1),m]+B \bigg )
e^{-i[\omega t-kx+\delta]}\,,
\ee
and
\be\label{6.50}
v(x,t)=D \sqrt{m} \cn[\beta(x-ct+\delta_1),m]\,,
\ee
is an exact solution to the coupled field equations (\ref{6.42}) provided
$A,D$ satisfy relations (\ref{6.47}) or the constraint (\ref{6.48}), 
while the other relations now are
\bea\label{6.51}
&&c = 2k= (2m-1)\beta^2\,,~~z = \frac{B}{A}=\frac{-(2-m)\pm \sqrt{1-m+m^2}}{3}
\,, \nonumber \\
&&\omega = k^2-2[1+m+3z]\beta^2-Ag[1-m+z]\,.
\eea

We now show that remarkably, the coupled model also admits even a linear 
superposition of the two as an exact solution, i.e. $\dn^2\pm \sqrt{m} \cn\dn$ as 
solution of the quadratic NLS field and $\dn\pm\sqrt{m}\cn$ as solution of
the MKdV field.   It is easily checked that 
\bea\label{6.52}
&&u(x,t)= \bigg (\frac{A}{2} \dn^2[\beta(x-ct+\delta_1),m] \nonumber \\
&&+ \frac{F}{2} \sqrt{m} \cn[\beta(x-ct+\delta_1),m] 
\dn[\beta(x-ct+\delta_1),m]+B \bigg )e^{-i[\omega t -kx +\delta]}\,,
\eea
and
\be\label{6.53}
v(x,t)=\frac{D}{2} \dn[\beta(x-ct+\delta_1),m]
+\frac{G}{2} \sqrt{m} \cn[\beta(x-ct+\delta_1),m]\,,
\ee
is an exact solution to the coupled field equations (\ref{6.42}) provided
\bea\label{6.54}
&&F=\pm A\,,~~G=\pm D\,,
~~z=\frac{B}{A}=-\frac{5-m \pm \sqrt{1+14m+m^2}}{12}\,, \nonumber \\
&&c = \frac{(1+m)}{2}\beta^2\,,~~\omega=k^2-\left[\frac{7+m}{2}+6z\right]
-gA\left[z+\frac{1-m}{4}\right]\,,
\eea
while $A,D$ satisfy the relations (\ref{6.47}) or the constraint 
(\ref{6.48}). Note that the signs of $F = \pm A$ and $G = \pm D$ are correlated.

\subsection{Quadratic NLS-NLS coupled Model}

Let us consider the following coupled QNLS-NLS model with field equations
\bea\label{6.55}
&&iu_t+u_{xx}+g_1|u|u + \alpha u |v|^2  =0 \, , \nonumber \\
&&iv_t+v_{xx} +g_2 |v|^2 v_{x}+\gamma v |u| =0\,, 
\eea
where $u$ and $v$ are the QNLS and the NLS fields, respectively. 
We first show that these coupled equations admit $\dn^2$ as solution of the 
quadratic NLS and either $\dn$ or $\cn$ as the solution of the NLS field. 

It is easily checked that 
\be\label{6.56}
u(x,t)= \bigg (A\dn^2[\beta(x-ct+\delta_1),m]+B \bigg )
e^{-i[\omega_1 t-kx+\delta]}\,,
\ee
and
\be\label{6.57}
v(x,t)=D \dn[\beta(x-ct+\delta_1),m] e^{-i[\omega_2 t -k_1 x+ \delta_2]}\,,
\ee
is an exact solution to the coupled field Eqs. (\ref{6.55}) provided
\be\label{6.58}
\alpha D^2+g_1 A =6\beta^2\,,~~g_2 D^2+\gamma A = 2\beta^2\,,
\ee
and further
\bea\label{6.59}
&&c = 2k= 2k_1\,,~~z = \frac{B}{A}=\frac{-(2-m)\pm \sqrt{1-m+m^2}}{3}
\,, \nonumber \\
&&\omega_1 = k^2-2[2(2-m)+3z]\beta^2-Ag_1 z\,,~~
\omega_2 = k^2-(2-m)\beta^2-A \gamma z\, . 
\eea
On solving Eqs. (194) we find that $A,D^2$ are given by 
\be\label{6.60}
A = \frac{2(\alpha-3g_2)\beta^2}{\alpha \gamma-g_1 g_2}\,,~~
 D^2 = \frac{2(3\gamma-g_1)\beta^2}{\alpha \gamma-g_1 g_2}\,.
\ee
However, in the special case when $3\gamma = g_1, \alpha = 3g_2$,
$A$ and $D$ are undetermined and instead of the relations (\ref{6.60}), 
we only have the constraint
\be\label{6.61}
3g_2 D^2+ g_1 A=6\beta^2\,.
\ee

It is easily checked that 
\be\label{6.62}
u(x,t)= \bigg (A\dn^2[\beta(x-ct+\delta_1),m]+B \bigg )
e^{-i[\omega_1 t-kx+\delta]}\,,
\ee
and
\be\label{6.63}
v(x,t)=D \sqrt{m} \cn[\beta(x-ct+\delta_1),m] 
e^{-i[\omega_2 t -k_1 x+ \delta_2]}\,,
\ee
is an exact solution to the coupled field equations (\ref{6.55}) provided
the relations (\ref{6.60}) are or the constraint (\ref{6.61}) is satisfied and 
further
\bea\label{6.64}
&&c = 2k= 2k_1\,,~~z = \frac{B}{A}=\frac{-(2-m)\pm \sqrt{1-m+m^2}}{3}
\,, \nonumber \\
&&\omega_1 = k^2-2(1+m+3z)\beta^2-Ag_1 (z+1-m)\,,~~
\omega_2 = k^2-(2m-1)\beta^2-A \gamma (z+1-m)\, . 
\eea

We now show that remarkably, the model also admits even a linear superposition 
of the two as exact solutions, i.e.$\dn^2 \pm \sqrt{m} \cn\dn$ as a solution of the 
quadratic NLS field and $\dn \pm \sqrt{m} \cn$ as the solution of the NLS field. 
It is easily checked that 
\bea\label{6.65}
&&u(x,t)= \bigg (\frac{A}{2} \dn^2[\beta(x-ct+\delta_1),m] \nonumber \\
&&+ \frac{F}{2} \sqrt{m} \cn[\beta(x-ct+\delta_1),m] 
\dn[\beta(x-ct+\delta_1),m]+B \bigg )e^{-i[\omega_1 t -kx +\delta]}\,,
\eea
and
\be\label{6.66}
v(x,t)=\bigg (\frac{D}{2} \dn[\beta(x-ct+\delta_1),m]
+\frac{G}{2} \sqrt{m} \cn[\beta(x-ct+\delta_1),m] \bigg )
e^{-i[\omega_2 t -k_1 x+\delta_2]}\,,
\ee
is an exact solution to the coupled field equations (\ref{6.55}) provided
\bea\label{6.67}
&&F=\pm A\,,~~G=\pm D\,,
~~z=\frac{B}{A}=-\frac{5-m \pm \sqrt{1+14m+m^2}}{12}\,, \nonumber \\
&&c = 2k = 2k_1\,,~~\omega_1=k^2-\left[\frac{7+m}{2}+6z\right]
-gA\left[z+\frac{1-m}{4}\right]\,, \nonumber \\
&&\omega_2 = k^2-\frac{1+m}{2}\beta^2-\gamma A\left(z+\frac{1-m}{4}\right)\,,
\eea
while $A,D$ satisfy the relations (\ref{6.60}) or the constraint 
(\ref{6.61}). Again note that the signs of $F = \pm A$ and $G = \pm D$ are
correlated.

\section{Summary and Conclusions}

In this paper we have demonstrated through many examples that a kind of linear 
superposition holds good in the case of several discrete as well as continuum 
nonlinear equations.  In particular, as a support to our conjecture, we have 
presented several nonlinear equations which admit $\dn(x,m)$ and $\cn(x,m)$ 
as periodic solutions, and have shown that all of them, without fail, also admit 
$\dn(x,m) \pm \sqrt{m} \cn(x,m)$ as periodic solutions. We would like to reemphasize  
that what we have proposed is only a kind of linear superposition and not the full 
linear superposition that is obtained in the linear theories. However, we find it 
remarkable that in spite of the nonlinear terms, even a kind of linear superposition 
seems to hold good in many nonlinear models, both discrete and continuum.  We
have also shown that such superposed solutions also exist in several {\it coupled} 
field theories. Many of these models such as MKdV, NLS, saturable discrete NLS, 
etc. have found wide ranging applications in many interesting physics problems 
\cite{1,2, ak4,of,bb}. While some of these models are integrable, many others are 
nonintegrable. It would be worth enquiring whether the new solutions we have 
obtained have specific physical relevance. Thus, it would be important to examine 
the (linear and nonlinear) stability of these newly found solutions. In this
context it is worth pointing out that the stability of the $\cn$ and $\dn$ 
solutions in the NLS and MKdV equations has been examined in detail in
recent years. Inspired by the earlier work of Rowlands \cite{row}, it has 
been shown that in the NLS case, the $\cn$ solution is 
unstable \cite{ivey}. However, in the MKdV case, it has been shown that the
$\dn$ solution is stable with respect to the periodic perturbations of the
same period and that this is true for all values of the modulus parameter
$m$. On the other hand, the stability of the $\cn$ solution changes as the
value of $m$ changes \cite{ang,niv}. Stability analysis has also
been done for the $\cn$ and $\dn$ solutions of the saturable discrete nonlinear
Schr\"odinger equation \cite{ak4} and it has been shown that both of these 
solutions are stable over a sizable amount of the parameter space.

Similarly we have also shown that several nonlinear equations which admit
$\dn^2(x,m)$ as a solution, also admit $\dn^2(x,m) \pm \sqrt{m} \cn(x,m) \dn(x,m)$ 
type superposed solution. While this cannot be regarded as a linear superposition,
since $\cn(x,m) \dn(x,m)$ is not a solution of these models, we find it rather 
remarkable that such solutions, without fail, hold good in several nonlinear 
equations. Further, we have also shown the existence of such solutions in 
many coupled systems which admit $\dn^2(x,m)$ as a solution in both the fields.
Since such solutions occur in physically important models like KdV, quadratic NLS, 
etc. \cite{1,2} it would be insightful to check the (linear as well as 
nonlinear) stability of such solutions. In this context it is worth pointing 
out that the stability of the $\dn^2$ solution has received some attention
in the literature. Inspired by the early works of Benjamin \cite{ben} and 
Bona \cite{bon}, It has been shown \cite{bol} that the $\dn^2$ solution is
stable with respect to perturbation of an arbitrary period or even with respect 
to perturbations that are quasi-periodic. As far as $\dn^2$ solution of the
$\phi^2$-$\phi^3$ field theory is concerned, one would expect it to be unstable.
Nevertheless, it would be worthwhile to explicitly check the stability of the $\dn^2$
as well as $\dn^2 \pm \sqrt{m}\cn \dn$ solutions in this case.   

In addition, we have also considered several coupled theories which admit $\cn(x,m)$ 
and $\dn(x,m)$ type solutions in one field and $\dn^2(x,m)$ type solutions in the other 
field and shown that these models also admit superposed solutions of the form $\dn(x,m) 
\pm \sqrt{m} \cn(x,m) \dn(x,m)$ in the first field and $\dn^2(x,m) \pm \sqrt{m} \cn(x,m) 
\dn(x,m)$ in the second field. Again, it is of physical importance to check the stability of 
these solutions. 

What could be the possible reason why a linear superposition of the form $\dn(x,m) \pm 
\sqrt{m} \cn(x,m)$ is a solution for several nonlinear equations which admit $\cn(x,m)$ 
and $\dn(x,m)$ as solutions? We believe that the main reason is that $\cn(x,m)$ and 
$\dn(x,m)$ functions are quite similar, both being an even function of its argument and 
both of them as well as their derivatives being identical at $m=1$.  This is in contrast to 
the $\sn(x,m)$ Jacobi elliptic function which is an odd function of its argument  and at 
$m=1$ it goes to $\tanh(x)$. This is in contrast to the even functions $\cn(x,m)$ and 
$\dn(x,m)$, both of which at $m=1$, go over to $\sech(x)$, and that is why $\cn(x,m) \pm 
\sn(x,m)$ superposition does not seem to work in the several examples that we have 
checked. 

Presumably there is a deeper reason to all this which needs to be explored. It would be 
worthwhile to find a rigorous proof of our conjecture. It would also be interesting if one 
can find a counter-example to our conjecture.  One line of thought is that the solutions 
are all of the traveling wave form and many of the PDEs can be reduced to an ODE 
either of the form $u+u_{xx}+u^2=c$ or $u+u_{xx}+u^3=0$.  Understanding these two 
simple ODEs might shed some light on the underlying general principle for the kind 
of superposition we have found.  Note that in both cases the superposed solutions exist;  
however, at present we are unable to draw any general conclusions about the proposed 
superposition.

\section{Acknowledgement} 

This work was supported in part by the U.S. Department of Energy. 

\section{Appendix: Dew Local Identities for Jacobi Elliptic Functions}

In this Appendix, we list the various local identities for Jacobi elliptic functions \cite{ak3} 
which have been used while deriving the various discrete solutions in Sec. IV.
\be\label{1.8}
\dn^2(x,m)[\dn(x+a,m)+\dn(x-a,m)]= 2\ns(a,m) \ds(a,m) \dn(x,m)-
\cs^2(a,m) [\dn(x+a,m)+\dn(x-a,m))]\,,
\ee
\be\label{1.9}
m\cn^2(x,m)[\cn(x+a,m)+\cn(x-a,m)]= 2\ns(a,m) \cs(a,m) \cn(x,m)-
\ds^2(a,m) [\cn(x+a,m)+\cn(x-a,m))]\,,
\ee
\be\label{1.10}
\dn^2(x,m)[\dn(x+a,m)-\dn(x-a,m)]= -2m\cs(a,m) \cn(x,m)\sn(x,m)-
\cs^2(a,m) [\dn(x+a,m)-\dn(x-a,m))]\,,
\ee
\be\label{1.11}
m\cn^2(x,m)[\cn(x+a,m)-\cn(x-a,m)]= -2\ds(a,m) \sn(x,m) \dn(x,m)-
\ds^2(a,m) [\cn(x+a,m)-\cn(x-a,m))]\,,
\ee
\bea\label{1.12}
&&\cn(x,m)\dn(x,m)[\dn(x+a,m)+\dn(x-a,m)]= \nonumber \\
&&2\ds(a,m) \ns(a,m)\cn(x,m)-
\cs(a,m) \ds(a,m) [\cn(x+a,m)+\cn(x-a,m)]\,,
\eea
\bea\label{1.13}
&&\cn(x,m)\dn(x,m)[\dn(x+a,m)-\dn(x-a,m)]= \nonumber \\
&&-2\cs(a,m) \sn(x,m)\dn(x,m)-
\cs(a,m) \ds(a,m) [\cn(x+a,m)-\cn(x-a,m)]\,,
\eea
\bea\label{1.14}
&&m\cn(x,m)\dn(x,m)[\cn(x+a,m)+\cn(x-a,m)]=  
2\cs(a,m) \ns(a,m)\dn(x,m) \nonumber \\
&&-\cs(a,m) \ds(a,m) [\dn(x+a,m)+\dn(x-a,m)]\,,
\eea
\bea\label{1.15}
&&m\cn(x,m)\dn(x,m)[\cn(x+a,m)-\cn(x-a,m)]= 
-2\ds(a,m) \sn(x,m)\cn(x,m) \nonumber \\
&&-\cs(a,m) \ds(a,m) [\dn(x+a,m)-\dn(x-a,m)]\,,
\eea


\end{document}